\documentstyle[epsf]{l-aa}
\begin{document}

  \thesaurus{02.          
              (08.14.1;   
               02.04.1;   
               02.13.1;   
               02.18.6)   
            }
\title{On cooling of magnetized neutron stars}
\author{Y.A.\,Shibanov and D.G.\,Yakovlev }
\institute{A.F.\,Ioffe Physical-Technical Institute,
           194021, St-Petersburg, Russia}

\date{Received 13 June 1995/ Accepted 25 September 1995}

\maketitle
\def\la{\;
\raise0.3ex\hbox{$<$\kern-0.75em\raise-1.1ex\hbox{$\sim$}}\; }
\def\ga{\;
\raise0.3ex\hbox{$>$\kern-0.75em\raise-1.1ex\hbox{$\sim$}}\; }
\newcommand{\om}{\mbox{$\omega$}}              
\newcommand{\th}{\mbox{$\vartheta$}}           
\newcommand{\ph}{\mbox{$\varphi$}}             
\newcommand{\ep}{\mbox{$\varepsilon$}}         
\newcommand{\ka}{\mbox{$\kappa$}}              
\newcommand{\dd}{\mbox{d}}                     
\newcommand{\vp}{\mbox{\boldmath $p$}}         
\newcommand{\vk}{\mbox{\boldmath $k$}}         
\newcommand{\vq}{\mbox{\boldmath $q$}}         
\newcommand{\vv}{\mbox{\boldmath $v$}}         
\newcommand{\kB}{\mbox{\boldmath $k_{\rm B}$}} 
\newcommand{\vect}[1]{\mbox{\boldmath $#1$}}   


\begin{abstract}
Cooling of neutron stars with dipole magnetic fields is simulated
using a realistic model of the anisotropic surface temperature
distribution produced by magnetic fields.
Suppression of the electron thermal conductivity
of outer stellar layers
across the field increases thermal isolation
of these layers near the magnetic
equator. Enhancement of the radiative and longitudinal electron thermal
conductivities in quantizing magnetic fields reduces thermal
isolation near the magnetic poles. The equatorial increase 
of the isolation is pronounced for $B \ga 10^{10}$ G,
while the polar decrease -- for $B \ga 10^{12}$ G.
The effects compensate partly each other, and the actual influence
of the magnetic fields on the cooling
is weaker
than predicted by the traditional theories
where the equatorial
effects have been neglected. 

          
\end{abstract}

\section{Introduction}                                        
Thermal X-ray radiation
has been observed recently with $ROSAT$
from at least four cooling neutron stars:
PSR 0833-45, PSR 1055-52,
PSR 0656+14 and Geminga (see, e.g., \"Ogelman 1995).
A soft ($T_{\rm s} \sim (5$ -- $10) \times 10^5$~K)
component of the observed spectra is most probably emitted from all the
visible neutron star (NS) surface. It
pulses with the pulsar period, and
the pulsed fraction, $\sim 10-30 \%$, varies with photon energy.
The spectra and the light curves of these middle-age
($10^4$ -- $10^6$ yr) NSs contain important information on
NS parameters and on the properties of superdense matter
in NS interiors.

The pulsations are most probably caused by
the temperature
variation over the NS surface. The natural cause
of the variation is non-uniformity of
the surface magnetic field and 
anisotropy of the heat transport in
strongly magnetized subphotospheric layers of
cooling NSs (e.g., Yakovlev \& Kaminker 1994).

So far most of the cooling calculations of NSs
(e.g., Nomoto \& Tsuruta  1987, Van Riper 1991) 
have been performed under the traditional
simplified assumption that the
magnetic field is radial everywhere over the stellar surface.
In this case the thermal energy is carried through
the subphotospheric layers
along the magnetic field lines, and the surface temperature
$T_{\rm s}$ is uniform and
determined by the longitudinal (along the field $\vect{B}$) thermal
conductivity. The latter conductivity is mainly enhanced
by the magnetic field, which increases 
$T_{\rm s}$ at given internal temperature $T_{\rm i}$. As a result,
the traditional cooling theories
predict (e.g., Van Riper 1991) that
the strong magnetic fields $B \ga 10^{12}$~G increase
$T_{\rm s}$ (and the photon luminosity) at the neutrino
cooling stage. At this stage, a NS is not
too old (age $t \la 10^4$--$10^5$ yrs)
and cools mainly via neutrino emission from the stellar interior.
On the other hand, strong fields decrease $T_{\rm s}$ and accelerate
the cooling at the subsequent photon cooling stage when the
neutrino emission becomes low and the star cools via
the thermal surface radiation.

The main disadvantage of the traditional approach is
that it neglects those parts of the NS
surface where $\vect{B}$ is essentially non-radial, and
the transverse thermal conductivity is important.
The aim of this paper is to use a realistic model
of heat transport and the related surface temperature distribution
valid for any magnetic field geometry in the NS surface layers.
The model generalizes the results
of Greenstein \& Hartke (1983), Van Riper (1989),
Schaaf (1990a), and Page (1995) who performed detailed
studies of the NS surface temperature either for restricted
magnetic field geometries or under specified assumptions
on thermal conductivity of stellar matter (see Sect.~2, for details).
Adopting this model
we will carry out the cooling calculations (Sect.~3). Our results show
(Sects.~4 and 5) that the effects of the magnetic fields on the NS cooling
are more sophisticated than anticipated previously.

\section{Surface temperature distribution}
Consider the temperature distribution over the surface of a
magnetized NS. Let the star be not too young ($t \ga$ 1 -- $10^3$ yrs),
so that the internal thermal relaxation is
achieved. Then the stellar interior is highly isothermal:
$\tilde{T}=T(r) \exp(\Phi(r))$ is constant within the star,
where $T(r)$ is the local internal temperature, $r$ is radial
coordinate, $\Phi(r)$ is the gravitational redshift function which
takes into account the effects of General Relativity
(e.g., Glen \& Sutherland 1980). The isothermal interior
is surrounded by a very thin
subphotospheric layer. Its thickness ($\la$ several
meters) is much smaller than the stellar radius
$R \sim 10$ km. This layer produces the thermal
isolation of the NS interior.
The temperature at the bottom of this layer is
$T_{\rm i}= \tilde{T} \exp(-\Phi_s)$, where $\Phi_s \approx
\Phi(R) = 0.5 \ln(1-R_g/R)$,
$R_g = 2 G M/c^2$ is the gravitational stellar radius,
and $M$ is the stellar mass.
While the star cools down, the isolating layer becomes thinner
(e.g., Gudmundsson et al. 1983, Yakovlev \& Kaminker 1994).
The layer consists of two sublayers. 
In the inner one, the electrons are degenerate,
and they are the main heat carriers.
In the outer sublayer, the electrons
are non-degenerate, and heat is mostly transported
by radiation.
In a rather cold NS, the surface layer may become
solid, and the non-degenerate sublayer disappears.

The effective surface temperature $T_{\rm s}$ is determined by the heat
transport through the isolating magnetized layer.
The problem of calculating $T_{\rm s}$ is complicated
(Yakovlev \& Kaminker 1994),
and the exact solution has not yet been found.
Let us discuss a realistic model
solution. Since the isolating layer is thin it is reasonable
to adopt a plane-parallel one-dimensional approximation almost
everywhere over the surface. We assume
that the magnetic field is constant in each local part
of the surface layer. Let
the thermal flux $F$ be normal (radial) to the surface
and the tangential flux be insignificant.
Then $F = \sigma T_{\rm s}^4$, where $T_{\rm s}$ is the local effective
surface temperature, and
$\sigma$ is the Stefan -- Boltzmann constant.
It is well known that $T_{\rm s}$
depends on the internal temperature $T_{\rm i}$,
and on the surface gravity $g_{\rm s} = (GM/R^2)/ \sqrt(1-R_g/R)$
($T_{\rm s} \propto g_{\rm s}^{1/4}$; see, e.g., Gudmundsson et al. 1983).
In our case, the local surface temperature
$T_{\rm s}=T_{\rm s}(B,\chi)$ depends also on the local magnetic field $B$
and on the angle $\chi$ between $\vect{B}$ and the normal to
the surface. If $T_{\rm s}(B,\chi)$ is known, the NS total
thermal luminosity is given by
\begin{equation}
       L(T_{\rm i}) =  \sigma R^2 \, \int T_{\rm s}^4(B,\chi) \, \dd \Omega
\label{eq:L}
\end{equation}
($\dd \Omega$ being solid angle of a surface element),
and the redshifted luminosity (for distant observers)
is $L^\infty = L (1-R_g/R)$.

Let the thermal energy in the isolating layer
be carried by the thermal conduction (no convection
or turbulent motion). Then the thermal balance yields
\begin{equation}
      (\kappa_\parallel \cos^2 \chi + \kappa_\perp \sin^2 \chi) \;
      {\dd T \over \dd z}= F,
\label{eq:Gradient}
\end{equation}
where $\kappa_\parallel$ and $\kappa_\perp$ are,
respectively, the longitudinal
and transverse (with respect to $\vect{B}$) thermal conductivities,
and $z$ is depth from the surface within the star in
the locally flat reference frame. The equation of thermal balance
should be supplemented by the equation of hydrostatic equilibrium;
we assume (as in all cooling theories)
that the magnetic field in the NS surface layers is force-free. 
Both equations should be integrated from the surface inside the
star yielding the temperature and density profiles in the
outer NS layer with given flux $F$ (i.e., with given
$T_{\rm s}$). With increasing $z$, the temperature $T(z)$
tends to a constant value to be defined as the internal temperature
$T_{\rm i}$. Combining the
solutions for different $T_{\rm s}$, $B$ and $\chi$
one can determine the dependence of the effective
local surface temperature $T_{\rm s}$ on $T_{\rm i}$, $B$ and $\chi$.

Since the thermal conductivities $\kappa_\parallel$
and $\kappa_\perp$ depend generally on $T$, $B$ and density $\rho$
the solution is complicated even
with the above simplifications. Instead,
we shall use the following model solution
(Greenstein \& Hartke 1983)
\begin{equation}
     T_{\rm s}^4(B,\chi) = T_\parallel^4(B) \cos^2 \chi +
     T_\perp^4(B) \sin^2 \chi,
\label{eq:Te}
\end{equation}
where $T_\parallel(B)=T_{\rm s}(B,0)$ is the surface
temperature for the case when
$\vect{B}$ is normal to the surface (at the magnetic pole), 
and $T_\perp(B)=T_{\rm s}(B,\pi/2)$ is the same for the
case when $\vect{B}$
is tangential to the surface (at the magnetic equator).

Let us discuss briefly the properties of Eq.~(\ref{eq:Te}).
First, (\ref{eq:Te}) yields exact solution
of the thermal conduction problem (\ref{eq:Gradient})
for $\vect{B}$ normal or tangential to the NS surface.
Second, (\ref{eq:Te})
reproduces the limit of $B=0$. Third, one can easily prove
that (\ref{eq:Te}) provides the
solution for the case
when $\kappa_\parallel \gg \kappa_\perp$. The case
is realized for not very weak magnetic fields and not too close
to the magnetic equator ($\kappa_\parallel \cos^2 \chi \gg
\kappa_\perp \sin^2 \chi$). This case is very important
since the transverse electron conductivity of degenerate
electrons is much smaller than the longitudinal one,
typically, at $B \ga 10^{10}$~G (e.g.,
Yakovlev \& Kaminker 1994). Fourth, Eq.~(\ref{eq:Te})
is exact for the case when $\kappa_\perp / \kappa_\parallel$
is constant
     (independent of depth $z$)
for a given local part of the NS surface layer.
The latter case is also important. For instance,
it is realized in a strongly
degenerate non-relativistic electron gas at not too high
fields $B$ if the electron conductivity is produced
by Coulomb scattering on atomic nuclei
and the nuclei constitute a gas or liquid
(e.g., Yakovlev \& Kaminker 1994). The model solution
(\ref{eq:Te}) can actually be inaccurate at high magnetic fields
near the magnetic equator where
the thermal flux tangential to the surface can be
substantial and the plane-parallel approximation can fail.
However these areas of the NS
surface give small contribution into the total stellar
luminosity (\ref{eq:L}), and, thus, they seem to be insignificant
for cooling simulations.

Using Eq.~(\ref{eq:Gradient})
one can obtain the following inequality of the thermal conduction
problem
\begin{equation}
     T_{\rm s}(B,\chi) \geq T_\parallel(B) \sqrt{ \cos \chi}.
\label{eq:Test}
\end{equation}
This inequality gives the lower limit to the effective
surface temperature
    as a function of $\chi$. It follows from Eq.~(\ref{eq:Gradient})
    by keeping the most important longitudinal
    thermal conductivity but neglecting the transverse
    conductivity (which is often much lower). In this case
    the surface temperature $T_{\rm s}(B,\chi)$ is naturally
    expressed through $T_\parallel(B)$.
If $\kappa_\parallel \gg \kappa_\perp$ and
$\kappa_\parallel \cos^2 \chi \gg \kappa_\perp \sin^2 \chi$
one can expect the real surface temperature to be very
close to the lower limit. The inequality
(\ref{eq:Te}) can be useful for testing numerical calculations
of $T_{\rm s}$ in magnetized NSs.

Equation (\ref{eq:Te}) was first 
obtained  by Greenstein \& Hartke (1983).
The authors derived (\ref{eq:Te}) as an estimate
assuming actually that
the thermal conductivities $\kappa_\parallel$
and $\kappa_\perp$ are constant in the isolating layer.
We emphasize that the equation is valid
under much wider assumptions. Recently the solution similar to
(\ref{eq:Te}) has been proposed by Page (1995) for a dipole
magnetic field. However, he
used $T_{\rm s}(B_{\rm p},0)$ instead of
$T_\parallel(B)$, and $T_{\rm s}(B_{\rm eq},\pi/2)$ instead of
$T_\perp(B)$ on the right-hand side ($B_{\rm p}$ and $B_{\rm eq}$
being the polar and equatorial
magnetic fields, respectively). The actual difference of his formula
from Eq.~(\ref{eq:Te}) is not large
but Eq.~(\ref{eq:Te}) is thought to be more adequate.

The advantage of Eq.~(\ref{eq:Te}) is that it reduces the
calculation of $T_{\rm s}(B,\chi$)
to solving two heat conduction problems: for $\vect{B}$
normal ($T_\parallel(B)$) and tangential ($T_\perp(B)$) 
to the surface, as if at the
magnetic pole and equator, respectively.
The longitudinal (polar) problem has been considered
very thoroughly by Van Riper (1989), and we 
use his results. We have taken the dependence
of $T_{\rm s}$ on $T_{\rm i}$ from his Fig.~29 for $B$= 0, $10^{11}$,
$10^{11.5}$, $10^{12}$, $10^{12.5}$, $10^{13}$,
$10^{13.5}$~G and for
the surface gravity $g_{\rm s} = 10^{14}$~cm~s$^{-2}$.
We have interpolated these data for intermediate values of $B$.
The transverse (equatorial) heat transfer problem has been
analyzed by Schaaf (1991a). We have taken the dependence of
$T_{\rm s}$ on $T_{\rm i}$ from his fitting Eqs.~(29) and Table~3 for
$B$= $10^{11}$, $10^{12}$, and $10^{13}$~G at
$g_{\rm s} = 10^{14}$~cm~s$^{-2}$. We have interpolated these data for
intermediate values of $B$, and extrapolated them to lower $B$
(if required) in such a way to reproduce the $B=0$ dependence
of Van Riper (1989). We have used the scaling relation
$T_{\rm s} \propto g_{\rm s}^{1/4}$ to extend the results to the
surface gravities of study (Sect.~3).

Equation~(\ref{eq:Te}) is valid for any magnetic field geometry.
For simplicity, we assume that the surface magnetic
field is {\it dipolar}.
Let $B_{\rm p}$ be the polar magnetic field,
and let us introduce the reference frame which is flat for a local
part of the NS subphotospheric layer.
Taking into account the effects of General Relativity
on the dipole field configuration
(Ginzburg \& Ozernoy 1964) one has
\begin{eqnarray}
   B_R & = & B_{\rm p} \,  \cos \Theta,
               \;\;\; B_\Theta = - B_{\rm p} \, f \,\sin \Theta,
\label{eq:B} \\
    f  & = & {(1-u) \ln(1-u) + u -0.5 u^2 \over
        \ln(1-u) + u + 0.5u^2},
\nonumber
\end{eqnarray}
where $B_R$ and $B_\Theta$ are, respectively, the radial and tangential
field components, $\Theta$ is the colatitude (with $\Theta=0$
at the pole), and
$u=R_g/R$. If $u \ll 1$
the General Relativity effects are negligible and $f=-1/2$.
With increasing $u$, the General Relativity effects
make the field more radial.

Figure 1 shows the surface temperature distributions
for a NS with $M=1.3\, M_\odot$ and $R=11.70$~km
for two internal temperatures: $T_{\rm i}=10^8$
and $10^7$~K, typical for the neutrino
and photon cooling stages, respectively, see Sects.~3 and 4 below.
The results can be explained as follows.
\begin{figure}
\begin{center}
\leavevmode
\epsfxsize=8.0cm 

\epsffile[60 60 535 510]{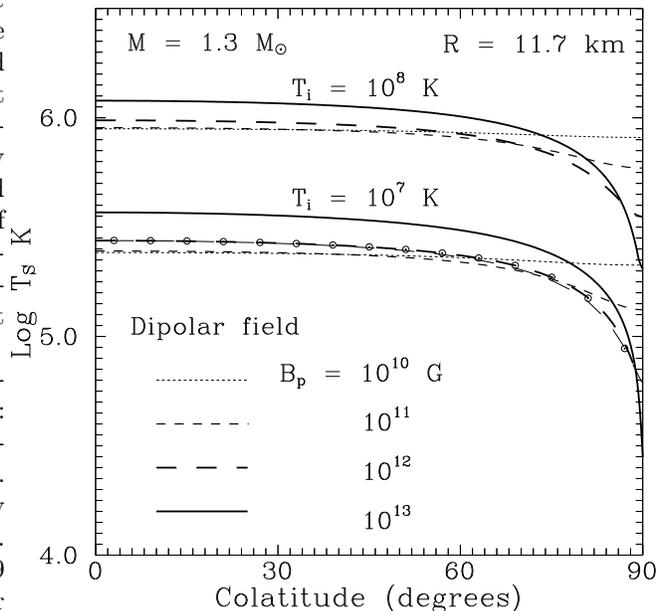}
\end{center}
\caption[ ]{
Distribution of the effective temperature $T_{\rm s}$
    from the pole (colatitude=0) to the equator (90)
of the 1.3\,$M_\odot$ neutron star
with the dipole magnetic field
for two values
of the internal temperature ($T_{\rm i}$ = 10$^7$ and 10$^8$~K)
and several polar field strengths $B_{\rm p}$. Open circles show
the lower $T_{\rm s}$ limit, Eq.~(\ref{eq:Test}),
for $B_{\rm p}=10^{12}$~G and $T_{\rm i}$ = 10$^7$~K.
Thin long dashes --- $T_{\rm s}$ for the same dipolar field
calculated without General Relativity effects
($f=-1/2$ in Eq.~(\ref{eq:B})).
}
\label{fig1}
\end{figure}
The magnetic fields affect the heat transfer in two
ways (e.g., Yakovlev \& Kaminker 1994).
First, the fields $B \ga 10^{10}$~G magnetize
the electrons in the degenerate sublayer of the outer 
isolating layer (see above).
Rapid electron Larmor rotation 
($\omega_B \tau \ga 1$, where $\omega_B$ is the
gyrofrequency of electrons at the Fermi surface, and $\tau$
is the electron relaxation time)
suppresses the transverse conductivity making 
the thermal conduction
anisotropic. Second, the fields
$B \ga 10^{12}$~G quantize electron motion --
the structure of the electron Landau levels become pronounced
in the thermodynamic and kinetic properties of
degenerate and non-degenerate electrons.

As seen from Fig.~1, the field $B_{\rm p} \leq 10^{10}$~G is too
low to modify noticeably the surface temperature $T_{\rm s}$ -- the
temperature
is nearly isotropic over the surface.
The magnetic fields $10^{10} \la B \la 10^{12}$~G
magnetize but not quantize
the electron motion. They
strongly reduce the transverse thermal conductivity of degenerate
electrons enhancing thermal isolation of the NS.
These fields do not affect significantly the longitudinal
electron conductivity, the radiative thermal conductivity
of outer non-degenerate sublayers, and thermodynamic
properties of matter. Accordingly the polar
surface temperature $T_\parallel$ remains nearly the same as for
$B=0$ but the equatorial temperature $T_\perp$
can be noticeably reduced.
With increasing $B$, the reduction
becomes stronger, and $T_{\rm s}(B,\chi)$ approaches the
minimum value $T_\parallel(B) \sqrt{ \cos \chi }$ given by
(\ref{eq:Test}) almost everywhere at the surface except
very near to the equator. As a result, the total luminosity
(\ref{eq:L})
       decreases
       with growing $B$
       but this decrease
saturates
after the inequality $\kappa_\parallel \gg \kappa_\perp$
is achieved.

The quantizing magnetic fields $B \ga 10^{12}$~G
induce strong quantum oscillations of the thermal conductivity
of degenerate electrons. Generally, these fields increase the
longitudinal electron conductivity. They also increase
the longitudinal and transverse radiative thermal
conductivity of non-degenerate layers.
Moreover, the strongly quantizing fields
reduce the pressure of degenerate electrons. All these
effects lower the thermal isolation.
Accordingly, they increase the
polar surface temperature
while the equatorial surface temperature remains strongly
damped by the magnetization of degenerate
electrons (Fig.~1, curves with $B_{\rm p} = 10^{13}$~G).
The net effect is to enhance
the total thermal luminosity $L$
of the star. With increasing $B$, the effect does not saturate
but becomes more pronounced.
The effects of the magnetic fields are stronger
for a colder NS, where the isolating layer is thinner and
more sensitive to the magnetic fields
(e.g., Kaminker \& Yakovlev 1994).

Therefore the thermal luminosity is suppressed by the magnetizing
fields in the equatorial region and enhanced
by the quantizing fields in the polar region.
Two effects, the suppression and the enhancement, act
in different ranges of $B$ and in opposite directions.
The traditional calculations of NS cooling assume that
the magnetic field is radial (polar) everywhere over
the surface. Accordingly the calculations
take into account the enhancement but neglect
the suppression. The model surface temperature
distribution (\ref{eq:Te}) allows one to include both
effects at once.

The validity of Eq.~(\ref{eq:Te}) can be checked by direct
2D simulations of the heat transport problem
in the surface layers of a magnetized NS. The first attempt to perform
2D simulations was made by Schaaf (1991b) under
many simplified assumptions.
Nevertheless, the 2D effects such as the heat flow from
the hotter polar regions to the cooler equatorial ones
produced by heat conduction or/and
meridional and convective motions should only
smooth the temperature variations
over the NS surface. In this respect $T_{\rm s}$ given by (\ref{eq:Te})
can be considered as an upper limit of the possible temperature
variation over the
NS surface. It seems sufficient for our cooling
calculations.

\section{Cooling calculations}
In order to analyze the effects of the magnetic fields on
the NS cooling we have performed
a series of numerical simulations.
We have used the cooling code of Gnedin \& Yakovlev (1993) and
Gnedin et al. (1994)
based on the approximation of isothermal NS interior.
The cooling consists of two stages: the neutrino
and photon ones. At the initial {\it neutrino} stage
($t \la (10^5$ -- $10^6)$ yrs) the NS cools mainly due to the 
emission
of neutrinos from its interior. The internal temperature
is ruled by the neutrino luminosity, and it is
independent of the surface temperature; the surface thermal
radiation just follows the internal cooling.
At the later {\it photon} stage the star becomes colder,
and the neutrino production mechanisms are inefficient.
The star cools mainly via thermal surface radiation which
governs the fall of the internal temperature.

In our simulations,
the equation of state of matter in the NS core
has been taken from Prakash et al. (1988). According to this
equation of state, the core consists of neutrons, protons
and electrons (no hyperons or exotic particles).
The maximum mass of a stable NS is 1.7\,$M_\odot$.
We have chosen two representative models, with $M$=1.3\,$M_\odot$ and
1.44\,$M_\odot$.

The model $M$=1.3\,$M_\odot$ ($R$=11.70 km, the central density
$\rho_c = 1.12 \times 10^{15}$ g~cm$^{-3}$) is typical for a
NS with the {\it standard} neutrino energy losses.
In the cores
of such stars, neutrinos are produced by the modified Urca processes
and the nucleon-nucleon neutrino-pair bremsstrahlung
(Friman \& Maxwell 1979, Yakovlev \& Levenfish 1995).
The modified Urca reactions are
$n+N \rightarrow p+e+N+\bar{\nu}_e$,
$p+e+N \rightarrow n+N+\nu_e$, where $N$ is a spectacular nucleon
required to satisfy momentum conservation.
Until recently, it has been widely assumed
that the main contribution into
the modified Urca comes from the neutron branch reaction
($N=n$). However according to Yakovlev \& Levenfish (1995)
the proton branch reaction ($N=p$) is of comparable efficiency, and
we have included it into the calculation. The nucleon-nucleon
bremsstrahlung ($N+N \rightarrow N+N + \nu + \bar{\nu}$)
consists of three branches ($n+n$, $n+p$, $p+p$), and all of
them have been incorporated using the results of
Friman \& Maxwell (1979) and Yakovlev \& Levenfish (1995).
The effective masses of neutrons
and protons (which enter the expressions for the neutrino energy
loss rates and heat capacities) 
have been set equal to 0.7 of the bare masses.

\begin{figure*}
\begin{center}
\leavevmode
\epsfxsize=11.8cm 
\epsffile[0 50 499 313]{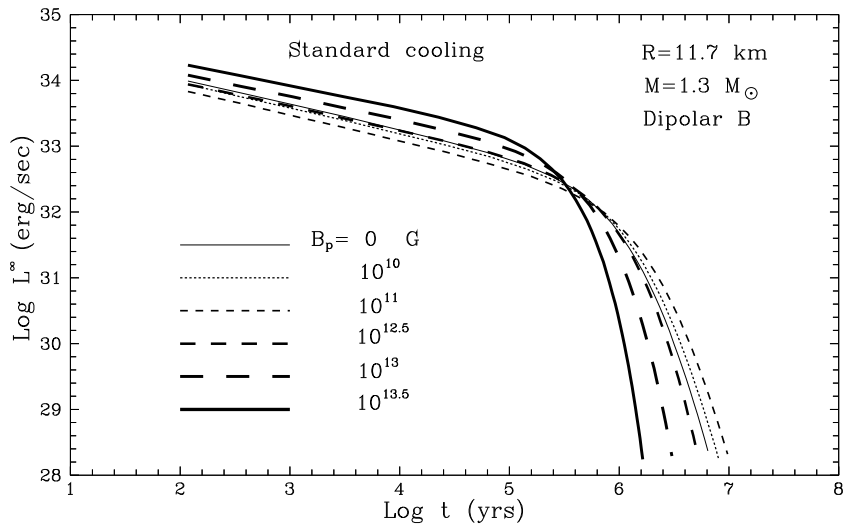}
\end{center}
\caption[ ]{
Photon thermal luminosity $L^\infty$
(redshifted for distant observers)
versus age of the 1.3\,$M_\odot$
neutron star (the standard cooling)
with a dipole magnetic field for several polar
field strengths $B_{\rm p}$.
}
\label{fig2}
\end{figure*}
%
\begin{figure*}
\begin{center}
\leavevmode
\epsfxsize=11.8cm

\epsffile[0 50 499 313]{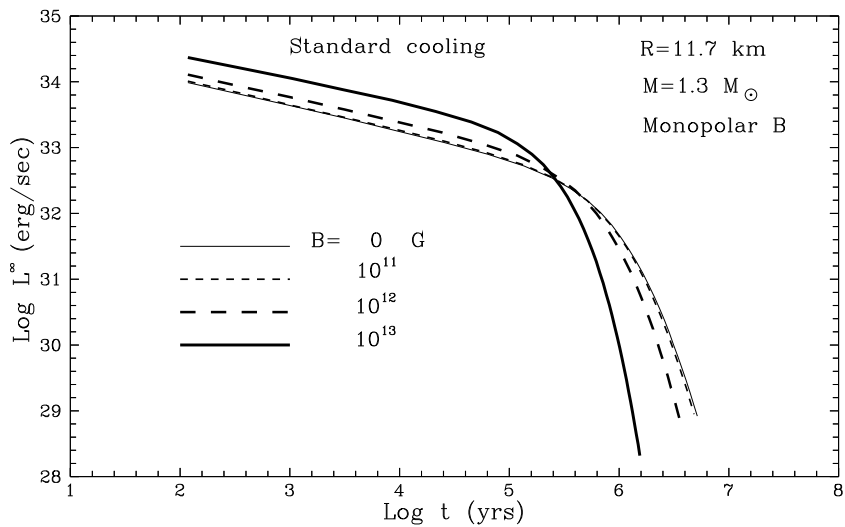}
\end{center}
\caption[ ]{
Same as for Fig.~2 but for radial (monopolar) magnetic
fields of several strengths $B$.
}
\label{fig3}
\end{figure*}

The model $M$=1.44\,$M_\odot$ ($R$=11.37 km,
$\rho_c = 1.37 \times 10^{15}$ g~cm$^{-3}$) represents a
massive NS whose neutrino luminosity is greatly
{\it enhanced} (Lattimer et al. 1991) by the direct Urca process
($n \rightarrow p+e+\bar{\nu}_e$, $p+e \rightarrow n+\nu_e$).
The latter process intensifies the
neutrino energy production rate by about 5 orders of magnitude for
$T_{\rm i} \sim 10^9$~K.
The direct Urca is a threshold reaction:
it is allowed (Lattimer et al. 1991) for certain
equations of state at rather high densities
(to satisfy momentum conservation of the reacting particles).
For the given equation of state, the critical
threshold density is $\rho_{cr} = 1.30 \times 10^{15}$ g~cm$^{-3}$.
In our NS model, the direct Urca is allowed in a small central kernel
of the star with $\rho > \rho_c$; the kernel radius is 2.32~km and
its mass is 0.035\,$M_\odot$. The neutrino luminosity of the kernel
greatly exceeds the neutrino luminosity from all other parts of
the star.

In both NS models we have also taken into account the neutrino
luminosity of the stellar crust produced by neutrino-pair
bremsstrahlung of electrons colliding with atomic nuclei.
This has been done in the same simplified manner as in
the work of Maxwell (1979). We have neglected
the contribution of the crust into the NS heat
capacity, since the crust mass is very low in our
  models. The mass of the crust in a NS with stiff
  equation of state and/or
  in a low-mass NS can be larger, i.e., in principle,
  the crust can contribute noticeably into the heat capacity.
  In our models,
  the heat capacity is produced by
neutrons, protons and electrons in the NS core.
We neglect possible superfluidity of neutrons and protons
since our main goal is to study the influence of the
surface magnetic fields on NS cooling. The latter influence
is determined by isolating properties of the NS
surface layers (Sect.~2), and it is not connected directly
with the internal superfluidity.

We have adopted the dipole magnetic field geometry,
Eq.~(\ref{eq:B}), as described in Sec.~2.
Using Eqs.~(\ref{eq:Te}) and (\ref{eq:B}) we can easily calculate
the photon thermal NS luminosity which is an important ingredient
of the cooling code. We have checked that the General Relativity
effects in the magnetic field distribution (\ref{eq:B})
are almost negligible  (Fig.~1) for our
NS models. For comparison,
we have also considered the case when the magnetic
field is radial and has the same strength everywhere over the
NS surface. This {\it monopolar} field geometry is
traditional for the NS cooling theories.
The field strengths have been varied from $B=0$ to
$B \sim 10^{13}$~G. We have followed the cooling until
the mean effective surface temperature drops below
$10^5$~K.
It has been unreasonable to proceed further
since the results of Van Riper (1989) and Schaaf (1991a)
used for calculating $T_{\rm s}$ are obtained
for a not too cold NS.

\section{Results and discussion}

Figures 2 and 3 show the standard ($1.3 \, M_\odot$)
cooling curves for the dipolar and monopolar magnetic fields
of different strengths. The curves display redshifted
NS photon luminosity versus age $t$.
   Since the surface temperature is distributed nonuniformly,
   the NS radiation,
   as detected by a distant observer,
   depends on angle between line of sight
   and the magnetic axis.
   We present the luminosity from the total NS surface
   which is important for NS cooling theories but does
   not necessarily determine the radiation flux in a specific direction.
The field geometry is seen to be very
important. The monopolar magnetic field always reduces
the NS thermal isolation (only the electron quantization
effects operate). Accordingly, the monopolar
field increases the photon luminosity $L$ at the neutrino
cooling stage ($t \la 3 \times 10^5$)~K, and decreases
$L$ at the subsequent photon cooling stage.
This is typical for the
traditional cooling calculations (e.g., Van Riper 1991).
On the contrary, the dipolar field $B_{\rm p}=10^{11}$~G, for example,
decreases $L$ at the neutrino cooling stage, and increases
at the photon stage. This is certainly because the magnetizing
fields enhance the equatorial thermal isolation.
However, for larger $B_{\rm p}$, the polar drop of the thermal
isolation becomes more important and produces just the
traditional cooling effect: it raises $L$ at the neutrino
cooling stage and decreases at the photon stage. With increasing $B_{\rm p}$,
the polar effect becomes stronger than the
equatorial one, but both effects interfere, and
the results are quantitatively different from the traditional
ones.

\begin{figure}
\begin{center}
\leavevmode
\epsfxsize=6.8cm 

\epsffile[60 60 520 520]{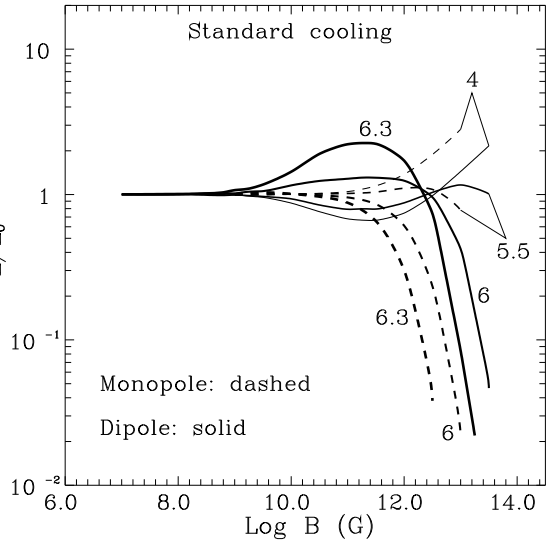}
\end{center}
\caption[ ]{
Ratio $L(B)/L(0)$ of the photon luminosities of magnetized and
non-magnetized 1.3\,$M_\odot$ neutron stars with
dipolar (Fig.~2) and monopolar (Fig.~3) fields versus
field strengths for several stellar ages $t$[yrs]
(values of $\lg (t)$ are given near the curves)
}
\label{fig4}
\end{figure}

The above effects are also illustrated in Fig.~4
where we plot $L(B)/L(0)$ versus $B$ for
the $1.3 \, M_\odot$ NS of several ages.
For the monopolar field, the dependence of
$L(B)/L(0)$ on $B$ at given age is monotonic, and one
needs $B \ga 10^{12}$~G to affect the cooling.
For the dipolar field, the same dependence is
non-monotonic, and the fields $B \ga 10^{10}$~G
affect noticeably the cooling curves.
It is remarkable that the dipolar fields
$B_{\rm p} \approx 3 \times 10^{12}$ almost do not affect
the NS cooling. For these fields, the enhancement
of the equatorial isolation nearly compensates
the drop of the polar one, and the cooling proceeds just
as if the NS were non-magnetized. Nevertheless the field
$B_{\rm p} \approx 3 \times 10^{12}$  makes the surface temperature
distribution very anisotropic (Fig.~1), which can
produce strong modulation of the surface thermal radiation
(Shibanov et al. 1995).
When $B_{\rm p} \la 3 \times 10^{12}$~G,
the equatorial effects prevail while for
$B_{\rm p} \ga 3 \times 10^{12}$~G the traditional polar effects
are stronger.
   Note that strong dependence of
   $L(B)/L(0)$ on $B$ for
   $\lg(t[yr])=$ 6 and 6.3
   is partially explained by steep slopes of the cooling curves
   (Fig.~2) at the photon cooling stage.
Figures 5 and 6 plot the cooling curves for the $1.44 \, M_\odot$
NS. The curves
for this (rapidly cooling) NS are qualitatively
the same as for the standard cooling (Figs.~2 and 3).
However the photon luminosity $L$ is much lower than for
the standard cooling since the rapid cooling goes much
faster at the neutrino cooling stage. The appropriate
luminosity ratios $L(B)/L(0)$ versus $B$ for several
stellar ages are shown in Fig.~7. The curves are
quite similar to those for the standard cooling (Fig.~4).
\begin{figure*}
\begin{center}
\leavevmode
\epsfxsize=12.0cm 
\epsffile[0 50 499 313]{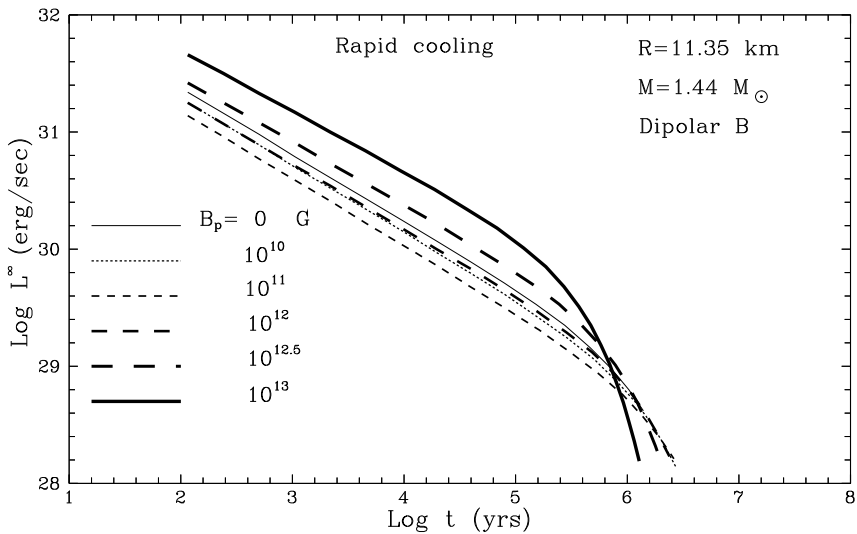}
\end{center}
\caption[ ]{
Same as in Fig.~2 for the 1.44\,$M_\odot$ neutron star
(rapid cooling).
}
\label{fig5}
\end{figure*}
%
\begin{figure*}
\begin{center}
\leavevmode
\epsfxsize=12.0cm 
\epsffile[0 50 499 313]{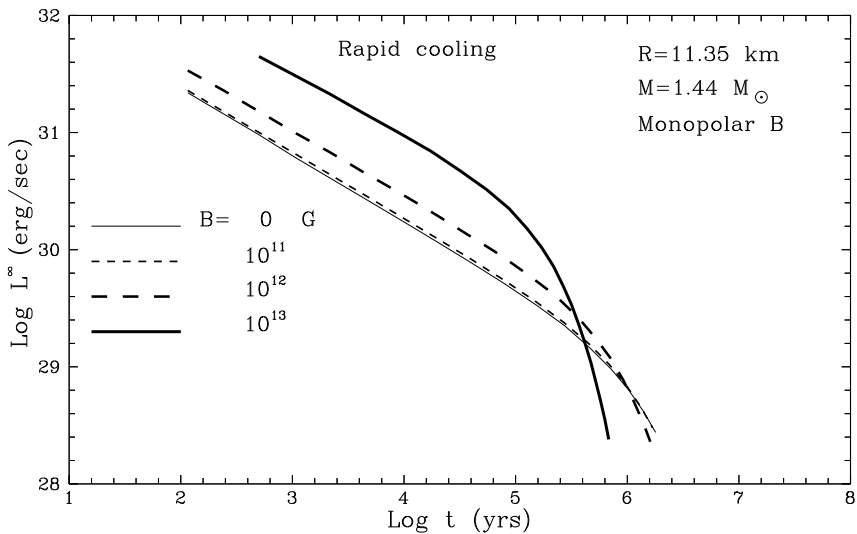}

\end{center}
\caption[ ]{
Same as in Fig.~3 for the 1.44\,$M_\odot$ neutron star.
}
\label{fig6}
\end{figure*}
%
\section{Conclusions}
We have used a realistic model (\ref{eq:Te})
of the anisotropic
distribution of the effective temperature $T_{\rm s}$ over the
surface of a magnetized NS and have calculated
NS cooling. We have considered both standard
and rapid cooling of the NS with the dipole magnetic field.
The results are noticeably different from the traditional
results obtained for radial (monopolar) magnetic fields.
We have shown (Figs.~2--7) that the dipole fields
$10^{10} \la B_{\rm p} \la 3 \times 10^{12}$~G
decrease the stellar photon
luminosity at the neutrino cooling stage
(by a factor of $\sim 2$), and increase
       the luminosity
at the photon cooling stage
(by about one order of magnitude),
    as compared to a non-magnetized NS,
contrary to
the traditional theories developed for monopolar
magnetic fields. The effect is produced by the
growth of the thermal isolation of the equatorial surface
layers due to the low thermal conductivity
of degenerate electrons across the magnetic field.
On the other hand, the dipole fields
$B_{\rm p} \ga 3 \times 10^{12}$~G increase the photon thermal luminosity
at the neutrino cooling stage and decrease
    the luminosity
at the photon cooling
stage,
   in comparison with the luminosity at $B=0$.
This conclusion agrees qualitatively with the results of
the traditional theories but the actual
magnetic field effect is much
weaker than predicted by these theories due to the interference
with the opposite effect
in the equatorial regions. We have obtained also that the dipole
field with $B_{\rm p} \approx 3 \times 10^{12}$~G has almost no
effect on the NS cooling.

Our results are based on a model of the surface temperature
distribution applied to the dipole magnetic field. It would be
easy to consider the NS cooling for other magnetic
field geometries (quadrupole, shifted dipole, etc.), and the
results are expected to be qualitatively the same as
for the dipole field: the parts of the surface with
essentially non-radial magnetic fields
should noticeably increase the thermal isolation (Sect.~2).

For further studies of the magnetic field effects
on the NS cooling,
it would be highly desirable to
perform detailed investigation of the heat transport 
in the outer NS layers using the best microscopic
physics available (equation of state, thermal conductivities, etc),
and obtain thus exact surface temperature distribution
in magnetized NSs. The
distribution (\ref{eq:Te}) used above is expected to give
an upper limit for the surface temperature variation
produced by the magnetic fields (Sect.~2). It can be used
for a test in more advanced theories. Even if (\ref{eq:Te})
is not very accurate at large magnetic fields near the
equatorial regions, it can yield quite accurate results
since the cold equatorial
surface parts do not contribute significantly into the
photon luminosity (\ref{eq:L}).
\begin{figure}
\begin{center}
\leavevmode
\epsfxsize=7.5cm
\epsffile[60 60 530 530]{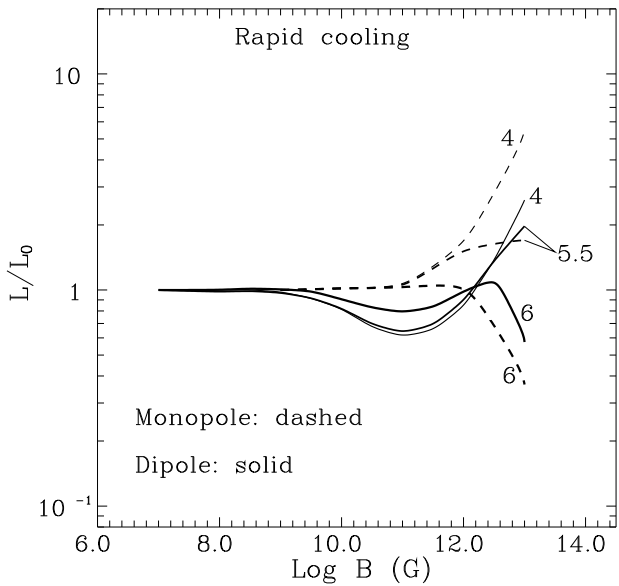} 

\end{center}
\caption[ ]{
Same as in Fig.~4 for the 1.44\,$M_\odot$ neutron star.
}
\label{fig7}
\end{figure}
The results of this work can be useful for
theoretical interpretation of thermal X-ray radiation of
NSs. All the objects from which this radiation has been
detected (Sect.~1) are thought to possess magnetic fields
of about several times of $10^{12}$~G. We can expect
that such fields do not affect significantly the NS
cooling but they can produce strongly anisotropic
surface temperature variation and associated
modulation of the surface thermal radiation.

The magnetic fields broaden the allowed
ranges of the photon luminosities in the $L$--$t$
diagram for NSs of various masses,
radii and equations of state with standard or
enhanced neutrino energy losses
(e.g., \"Ogelman 1994). According to the traditional theories,
high magnetic fields {\it increase} the upper boundary of
the expected photon luminosity at the neutrino cooling stage
(typically, at $10^2 \la t \la 3 \times 10^5$~yrs) and decrease
the lower boundary of $L$ at the photon stage.
Our results (Figs.~4 and 7) show that
the dipolar fields $B \sim 10^{11}$~G
{\it  decrease} the lower boundary of $L$ at the neutrino stage
and increase the upper boundary at the photon stage.
                                            
\acknowledgements 
We are grateful to Oleg Gnedin and Kseniya Levenfish for
the assistance with the cooling code.
We are also thankful to George Pavlov and Dany Page
for stimulating discussions,
    and to anonymous referee
    for many critical comments.
This work was partly supported
by RBRF, grant 93-02-2916,
ISF, grant R6A-000,
ESO C\&EE Programme, grant A-01-068, and INTAS, grant 94-3834.

\end{document}